\documentclass[11pt]{article}
\usepackage{array,epsfig,amsmath,amssymb}
\renewcommand{\baselinestretch}{1}
\makeatletter
\renewcommand{\section}{\setcounter{equation}{0}\@startsection
  {section}%
  {1}%
  {0pt}%
  {-1\baselineskip}%
  {0.4\baselineskip}%
  {\bfseries}}%
\renewcommand{\subsection}{\@startsection
  {subsection}%
  {2}%
  {0pt}%
  {-0.75\baselineskip}%
  {0.2\baselineskip}%
  {\bfseries}}%
\renewcommand{\subsubsection}{\@startsection
  {subsubsection}%
  {3}%
  {0pt}%
  {-0.5\baselineskip}%
  {0.1\baselineskip}%
  {\sc}}%
\makeatother

  {\end{list}}%

\def\De{\Delta}

\def\la{\lambda}        
\def\La{\Lambda}

\def\r{\rho}
\def\si{\sigma}

\def\eps{\epsilon}
\def\var{\varepsilon}
\def\om{\omega}

\def\tQ{{Q}}

\def\ds{\displaystyle}

\def\to{\rightarrow}
\def\l{\left}
\def\r{\right}

\def\bsigma{\boldsymbol{\rm \sigma}}
\def\nn{\nonumber}
\def\tn{\textnormal}
\def\ra{\rangle}

\newtheorem{theorem}{Theorem}[section]
\newtheorem{lemma}[theorem]{Lemma}

\newenvironment{proof}[1][Proof]{\begin{trivlist}
\item[\hskip \labelsep {\bfseries #1}]}{\end{trivlist}}

\begin{document}
\begin{titlepage}
\rightline{UCM-FTI/05-43}

\vskip 1.5 true cm
\begin{center}
{\Large \bf States with $v_1=\lambda, v_2=-\lambda$ 
and reciprocal equations in the six-vertex model
\\[9pt]}
\vskip 1.2 true cm 
{\rm M.J.~Rodr\'{\i}guez-Plaza}\footnote{E-mail: mjrplaza@fis.ucm.es}\\ 
\vskip 0.3 true cm
{\it Departamento de F\'{\i}sica Te\'orica I, 
     Facultad de Ciencias F\'{\i}sicas}\\ 
{\it Universidad Complutense de Madrid, 28040 Madrid, Spain}\\
\vskip 1.2 true cm

{\leftskip=25pt \rightskip=25pt 
\noindent

Abstract 

The eigenvalues of the transfer matrix in a six-vertex model (with
periodic boundary conditions) can be written in terms of $n$ constants
$v_1,\ldots,v_n$, the zeros of the function $Q(v)$.  A peculiar class
of eigenvalues are those in which two of the constants $v_1,v_2$ are
equal to $\lambda, -\lambda$, with $\Delta=-\cosh\lambda$ and $\Delta$
related to the Boltzmann weights of the six-vertex model by the usual
combination $\Delta=(a^2+b^2-c^2)/2 a b$.  
The eigenvectors associated to these eigenvalues 
are Bethe states (although they seem not).  We count the number of
such states (eigenvectors) for $n=2,3,4,5$ when $N$, 
the columns in a row of a square lattice, is arbitrary.
The number obtained is
independent of the value of $\Delta$, but depends on $N$.  
We give the explicit expression of the
eigenvalues in terms of $a,b,c$ (when possible) or in terms
of the roots of a certain reciprocal polynomial, being very
simple to reproduce numerically these special eigenvalues for 
arbitrary $N$ in the blocks $n$ considered.
For real $a,b,c$ such eigenvalues are real.
%The eigenvalues of the transfer matrix in the square six-vertex model
%(with periodic boundary conditions) are given in terms of $n$ constants
%$v_1,\ldots,v_n$. In this paper we study the eigenvalues and eigenvectors
%such that two of those constants, $v_1$ and $v_2$, are equal to 
%$v_1=\lambda, v_2=-\lambda$, where $\lambda$ is related to the
%Boltzmann weights of the vertex $a,b,c$ through the combination
%eigenvalues characterized by  
\par}
\end{center}

%\vfill

\section{The problem}
\label{theproblem}

Some time ago the author of this note read in the paper {\it
Completeness of the Bethe Ansatz for the Six and Eight-Vertex Models}
by R.J~Baxter \cite[Sect.~4]{B1} the following sentence concerning
certain proper states of the transfer matrix in the 
six-vertex model at zero-field:
%(since we dont know a proper name for them we 
%will refer to them as {\em bound pair states}
%of the six-vertex model at zero-field):
\end{titlepage}
\setcounter{page}{2}

%----------------------------------------------------- Paper

{\leftskip=8pt \rightskip=8pt 

{\small The other problem that we encountered first occurs for $N=4$
and $n=2$, then for even $N$ and $2\le n\le N-2$. It is referred to by
Bethe himself and has been 
considered by others
since\footnote{\cite[after eq. (23)]{B},\cite{Sid}, \cite{NLK}, \cite{Ba}}. 
For some eigenvalues
with momentum $\pm 1$,
\footnote{Baxter means 
$e^{\displaystyle{i(k_1+\cdots+k_n)}}=\pm 1$}
i.e. $k_1+\cdots+k_n=0$ or $\pi$, we found that 

\vspace{-12pt}
\begin{equation*}
Q(v)=\displaystyle{\prod_{j=1}^n}\sinh\,[(v-v_j)/2]
\end{equation*}

\vspace{-12pt}
\noindent
had a pair of zeros $v_1, v_2$ such that $v_1=\lambda, v_2=-\lambda$.
\par
}}

\noindent
The lines continued later as follows:

{\leftskip=8pt \rightskip=8pt 
\noindent  
{\small For $N=4$ there was just one such eigenvalue $\Lambda$, in the
$n=2$ central block. For $N=6$ there was one in the $n=2$ block, two in
the $n=3$ block, and one in the $n=4$ block. For $N=8$ there were
$1,2,5,2,1$ in the $n=2,3,4,5,6$ blocks, respectively. This suggests
(tentatively) that the Catalan numbers may count such 
eigenvalues.\footnote{Catalan numbers are $1,2,5,14,132,429,\ldots$}  
The
momenta were $-1$ except for a single eigenvalue with momentum $+1$ in
each block with $3\le n\le N-3$.}
\par
}

If the author had understood properly the eigenvectors associated to
such eigenvalues and how to obtain them from Bethe ansatz, probably
would have not detained so long 
when reading these sentences. But that was not
the case: we were calculating at that time the free-energy per site of
a vertex model whose ground state was a state of this type,
and the value of the free-energy that we were deriving was once and
again the incorrect one. We decided in consequence to put aside the
free-energy problem for a time and study instead these
states in the six-vertex model.  We ignore the correct 
name that we shall use for them.
In the literature they have
received the name of {\em singular} Bethe states or {\em singularities}
of the Bethe solutions \cite{Sid,NLK,BMSZ},
and also {\em non-Bethe}
eigenvectors \cite{FKT}. We might even remember some references in
which they are alluded as {\em improper} states. Since they need a 
name and no other states are considered in this paper we will 
refer to them as {\em bound pairs} merely. 

This note communicates some results of the study and answers the
interrogation suggested in Baxter's paper: 
{\it Are Catalan numbers counting the bound pair states of a
square six-vertex model with periodic boundary conditions?}

\section{The model}
\label{model}
 
The model to be considered is a six-vertex model in a square
lattice \cite{Lieb, Baxterbook}. In this model to each 
site of the lattice
is associated one of the six arrangements of arrows shown in figure~1, 
where each of these arrangements has an energy 
$\var_1,\ldots,\var_6$ and a Boltzmann weight given by
\[
\om_j={\rm exp}(-\var_j/k_B\,T),\qquad j=1,\dots,6.
\]
The configurations of arrows satisfy the `ice rule', because at each site of
the lattice there are two arrows in and two arrows out.
  
\vspace{8pt}
\centerline{\epsfig{file=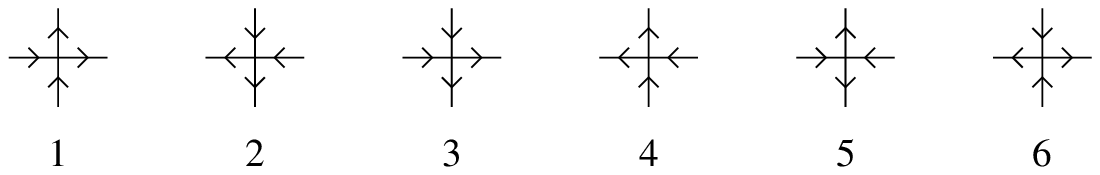,width=0.70\textwidth}}
\noindent\textsf{\footnotesize Figure~1.
The six configurations allowed at a vertex.  At each site of
the lattice there are two arrows in and two arrows out. This is known
as the `ice-rule'.}

\noindent
Suppose that the lattice has dimensions $M\times N$, that is $N$ sites
horizontally and $M$ vertically,  with the
imposition of periodic boundary conditions in both directions. 
The state of an arbitrary row of $N$
vertical edges is then specified by the configuration of up and down
arrows on the edge. Let $\bsigma=(\si_1,\ldots,\si_N)$ denote the
state ($\si_j=+1$ for an up arrow at vertex $j$, $\si_j=-1$ for a down
arrow).  If $\bsigma$ is the state of a row and $\bsigma'$
the state of the row bellow, the two adjacent states are coupled by
the transfer matrix $T_{\sigma\sigma'}$, whose entries are 
given by a trace of 
$2\times 2$ matrices  
\begin{equation}
T_{\sigma\sigma'}=\tn{ trace}\, R_{{\sigma_1\sigma'_1}}
\,R_{\sigma_2\sigma'_2}\cdots R_{\sigma_N\sigma'_N},
\label{transfermatrix}
\end{equation}
where
\[
R_{++}=\begin{pmatrix}\om_1 &0\\0&\om_4\end{pmatrix},\quad
R_{+-}=\begin{pmatrix} 0&\om_5\\0&0\end{pmatrix},\quad
R_{-+}=\begin{pmatrix} \om_6&0\\0&0\end{pmatrix},\quad
R_{--}=\begin{pmatrix} \om_3&0\\0&\om_2\end{pmatrix}.
\]
A consequence of the `ice rule' together with the horizontal
periodicity of the lattice is that the number $n$ 
of down (or up) arrows in a row is a conserved quantity from row to row, 
and $T$, a $2^N\times 2^N$ matrix, breaks up into $N+1$
diagonal blocks 
with one block for each value $n=0,1,\ldots,N$. The dimension of 
block $n$ is $N\choose n$.
The transfer matrix is used 
to calculate in statistical mechanics
the partition function of the lattice 
$Z=\tn{ trace}\,T^M$,
and this has implied the diagozalization of matrix $T$. 
In the case of a zero electrical field (the case treated here) 
where
\begin{equation}
a=\om_1=\om_2,\quad b=\om_3=\om_4,\quad c=\om_5=\om_6,
\label{zerofield} 
\end{equation} 
the eigenvalues $\Lambda$ of the transfer matrix are known 
to be \cite{Baxterbook}
\begin{equation}
\La(v)=(-1)^n\,\frac{\phi(\la-v)\,Q(v+2\la)+\phi(\la+v)\,Q(v-2\la)}{q(v)},
\label{Lambda1(v)}
\end{equation}
where functions $\phi(v)$, $Q(v)$ are
\begin{align}
\phi(v)&=\rho^N{\sinh^N\bigl(v/2\bigr)}
\label{defphi}\\
Q(v)&=\displaystyle{\prod_{j=1}^n}\sinh\,[(v-v_j)/2],
\label{defq}
\end{align} 
and $\rho$, $\la$, $v$ are defined so that
\begin{equation}
a=\rho\, \sinh\frac{1}{2}\,(\la-v),\quad 
b=\rho\, \sinh\frac{1}{2}\,(\la+v),\quad 
c=\rho\, \sinh\la.
\label{vlambda}
\end{equation}
To write the eigenvalues (\ref{Lambda1(v)}) we have to locate 
$v_1,\ldots,v_n$ for these eigenvalues. There are many solutions, 
corresponding to the different eigenvalues.

\section{Catalan numbers or not}
\label{Catalan}

{\it Do Catalan numbers count the bound pair states of a
square six-vertex model with periodic boundary conditions?} 
The answer is {\it no}.

\noindent
As was communicated in ref.~\cite{B1}, it is true that for $N=4$ 
there is one
bound pair in the $n=2$ central block, that for $N=6$ there are
$1,2,1$ in the $n=2,3,4$ blocks, respectively, and that for $N=8$
there are $1,2,5,2,1$ in the $n=2,3,4,5,6$ blocks. However, if
the counting started in the previous reference 
had continued it would have found that there are $1,2,6,10$ in
$n=2,3,4,5$\footnote{We omit to mention the blocks $n=N/2+1$ to
$N-2$ since the number of these states is the same as in the blocks
$2,\ldots,N/2-1$ but in reverse order} for $N=10$, and $1,2,7,12$ in
$n=2,3,4,5$ for $N=12$.
In fact our calculations here show that  
for general even $N$ the number 
is exactly $1,2,N/2+1,N$ in the blocks
$n=2,3,4,5$. The numbers
of states for $n=6$ and beyond wont be studied in this paper.
% \footnote{We omit the number of states 
% beyond $n=N/2$ because there
% are the same number that for $2 \le n<N/2$ but in the opposite order}

\section{Bound pairs and Bethe Ansatz}
\label{BA}

To obtain the eigenfunctions of the transfer matrix 
one can either diagonalize exactly the matrix (impossible when the size
is not reasonable) or use the Bethe ansatz, the trial
form that Bethe used for diagonalizing the quantum-mechanical
Hamiltonial of the one-dimensional Heisenberg model \cite{B}. 
The ansatz suggests 
that
the eigenstate of $T(v)$, 
$T(v)|\psi\rangle=\La(v)|\psi\rangle$,
 can be written as $|\psi\rangle=
\sum_{x_1<\ldots < x_n} f(x_1,\ldots,x_n)\,|x_1,\ldots, x_n\rangle$,
where the coefficients $f(x_1,\ldots,x_n)$ are 
\begin{equation}
f(x_1,\ldots,x_n)=\sum_{P} A_{p_1,\ldots,p_n}
\,e^{i k_{p_1} x_1}\cdots\,e^{i k_{p_n} x_n}.
\label{Bethestate}
\end{equation}
The numbers $x_1,\ldots, x_n$ indicate the positions of
$n$ down arrows
on the lower vertical edges of a row of the lattice,
and are ordered so that $1\le x_1< x_2 < \ldots x_n\le N$.
We have experienced that 
the coefficients $A_{p_1,\ldots,p_n}$ suitable to construct 
bound pairs\footnote{Because they give the same result that 
when the transfer matrix is directly diagonalized. 
When the size of the matrix allows it it is possible  
to carry many numerical experiments 
and they confirm this choice for the amplitudes}
are given by
\begin{equation}
A_{p_1,\ldots,p_n}=\eps_P/C \prod_{1\le i<j\le n}\, s_{p_j,p_i},
\label{As}
\end{equation}
where $\eps_P=\pm 1$ is the sign of the permutation $\{p_1,\ldots,p_n\}$ 
of $\{1,\ldots,n\}$ and $C$ is a non-zero constant to be fixed later 
in the most convenient manner (usually normalization). 
The vertex model defined by activities $a,b,c$
so that
\begin{equation}
\Delta=\frac{a^2+b^2-c^2}{2\,a b}
\label{defDelta1}
\end{equation}
enters in $s_{ij}$, 
defined as
\begin{equation}
s_{ij}=1-2\,\De \,e^{i k_j}+e^{i (k_i+k_j)}.
\label{sij}
\end{equation}
To write the eigenstates is only necessary then to 
know the factors
$e^{i k_1},\ldots,e^{i k_n}$ that appear in (\ref{Bethestate})
and (\ref{sij}). These factors are the solutions of the equations
\begin{equation}
e^{\displaystyle{i N k_{p_1}}} A_{p_2,\ldots,p_n,p_1}
=A_{p_1,\ldots,p_n},
\label{amplitude}
\end{equation}
that impose the periodic boundary conditions 
on the problem making that
$f(x_1,x_2,\ldots,x_N)=f(x_2,\ldots,x_N,x_1+N)$ what 
identifies the $N+1$ and $1$ vertices.
%that take into account the behaviour of the row 
%under the change of configuration from $|x_1,\ldots, x_n\rangle$ to
%$|x_1+1,\ldots, x_n+1\rangle$,
%\footnote{I.e., when each vertex is shifted to the next vertex on the right}
%identifying the $N+1$ and $1$ vertices. 
To be (\ref{amplitude})
consistent equations among
themselves, it is necessary that
\begin{equation}
e^{\displaystyle{i N (k_1+\cdots+k_n)}}=1.
\label{one}
\end{equation}
Equations 
(\ref{Bethestate}), (\ref{As}), (\ref{sij}),
(\ref{amplitude}) and (\ref{one}), are {\em sufficient} equations  
to write bound pair eigenfunctions, and when needed 
we will refer to them as
``the Bethe ansatz equations for bound pairs''. 
However, and this is not less important, 
it is also necessary 
a {\em correct normalization} of the
eigenfunction. Without it, the state cannot be obtained.
We have learned the correct normalization in
ref. \cite[Sect.~4]{B1}, and show an example later 
for $N=6$ and $n=3$. Equations 
(\ref{ba4-1}), (\ref{ba4-2}) and (\ref{ba4-1}), (\ref{ba4-2}), 
(\ref{ba5-3})
are deduced taking into account 
such normalization.

It is important to write, before finishing, the relation between 
$k_1\ldots,k_n$  
and $v_1,\ldots,v_n$ in (\ref{defq}) 
(or better between $e^{i k_j}$ and $e^{v_j}$)
\begin{equation}
e^{i k_j}=\frac{e^\la-e^{v_j}}{e^{\la+v_j}-1},
\label{eikj}
\end{equation}
as mentioned in many papers. This relation permits to move from the
eigenvalue (\ref{Lambda1(v)})
to the eigenvector (\ref{Bethestate}) of the transfer matrix 
when we precise it. 
%
%\begin{equation}
%s_{p_j,p_{j+1}} A_{p_1,\ldots,p_n}+s_{p_{j+1},p_j} 
%A_{p_1,\ldots,p_{j+1},p_j,\ldots,p_n}=0
%\label{sAAs}
%\end{equation}
%% (n-1)n! equations. But each equation appears twice: then
%% (n-1)n!/2 different equations

\section{A change of variables}
\label{change}

Before describing any eigenvalue we 
make a useful change of variables concerning $v$ and 
$\lambda$ in (\ref{vlambda}). The change is convenient for those
(the author in this specific problem among them) who
prefer to work
with polynomials rather than with 
hyperbolic functions as in (\ref{defq}). Define 
the variables 
\begin{equation}
z=e^{-v}, \qquad y= e^{-\lambda}
\label{defzy}
\end{equation}
instead of $v$ and $\lambda$, then (\ref{defq}) is
essentially\footnote{Essentially means up to multiplicative constants
that do not depend on $z$ (they may depend on $y$ because $y$
is regarded as a constant: after all $y$ is fixed by the value that we
choose for $\Delta$, and viceversa).  
The constants are not relevant because do not
change the value of $\Lambda(v)$, as commented after equation
(\ref{functionalrelation})} the polynomial in $z$ and $1/z$ given by
\begin{equation}
\tQ(z)=\frac{1}{z^{n/2}}\displaystyle{\prod_{j=1}^n}(z-z_j),
\label{definitionQ}
\end{equation}
with $z_j=e^{-v_j}$, $j=1\ldots,n$. To be correct we should have 
defined 
another symbol for (\ref{definitionQ}), $\tilde Q(z)$ for instance, 
however we will use the same letter with the understanding that 
$Q(v)$ stands for (\ref{defphi}) and $\tQ(z)$ for (\ref{definitionQ}).
In terms of these variables and together with
definitions 
(\ref{defphi})
and (\ref{defq}), relation (\ref{Lambda1(v)}) becomes
\begin{equation}
(2/\rho)^N \Lambda(v)\,Q(z)
=\frac{(-1)^n}{(zy)^{N/2}}\,\l[\l(z-y\r)^N Q(zy^2)+\l(1-zy\r)^N Q(z/y^2)\r],
\label{functionalrelation}
\end{equation}
where multiplicative constant factors in $\tQ$
cancel out of the calculations. 
To operate in a computer we prefer to work with this relation 
more than with (\ref{Lambda1(v)}).

\section{n=2}
\label{n=2}

This is the simplest case to study because the transfer matrix of $N$
edges (with $N$ even)
has
only one bound pair state in this block for 
arbitrary $\Delta$ defined in (\ref{defDelta1}). 
Since bound pairs are characterized by 
$v_1=\lambda$, $v_2=-\lambda$ as mentioned in Sec.~\ref{theproblem}, 
function (\ref{definitionQ}) factorizes as 
\begin{equation}
\tQ(z)=(z\,y-1)(z-y)/z,
\label{Qn=2}
\end{equation}
the zeros of $\tQ(z)$ being $z_1=y$ and $z_2=1/y$. Introduced this
function in (\ref{functionalrelation}) and noting
that the r.h.s. is {\em exactly} divided by 
$\tQ(z)$ in the l.h.s, the quotient affords the eigenvalue\footnote{
$\Lambda(v)$ in (\ref{functionalrelation}) is obtained in terms
of $z$ and $y$, of course. We have reexpressed the result in terms 
of $a,b,c$ to write (\ref{LambdaNn=2})}
\begin{equation}
\Lambda(v)=a^2 b^2\,(a^{N-4}+b^{N-4})-c^2\,(a^{N-2}+b^{N-2}),\qquad N\ge 4,
\quad n=2,
\label{LambdaNn=2}
\end{equation}
that is valid for generic $N$ even. It can be checked numerically that
(\ref{LambdaNn=2})
is always an eigenvalue of the transfer matrix for all
values of $a,b,c$ real or complex,\footnote{$a,b,c$, the Boltzmann weights 
(\ref{zerofield})
of the vertex model, are real and positive, but when 
diagonalization of a matrix is considered in general, with
no restriction to physical values only, they can also 
be negative or complex} 
and since the block $n=2$ is among the blocks of smallest
dimensions, it can be done even for $N$ not too small.
% considerable important/big
The eigenvector associated to (\ref{LambdaNn=2}) was known to Bethe
himself \cite[also after eq. (23)]{B} and is proportional to
\begin{equation}
|\psi\rangle=\sum_{l=1}^{N}(-1)^l\,|l,l+1\ra,
\label{knowntoBethe}
\end{equation}
after appropriate normalization. We do not reproduce here this 
eigenvector with the  Bethe ansatz (the example that we reproduce
is for $N=6$, $n=3$ later), but want to comment 
about $e^{i k_1}$ and   
$e^{i k_2}$. The product of these two factors is for the eigenfunction 
(\ref{knowntoBethe}) equal to
$-1$, since from (\ref{Bethestate}) derives the relation
\begin{equation}
f(x_1+1,x_2+1)=e^{\displaystyle{i(k_1+k_2)}}f(x_1,x_2),\quad 
\tn{with}\quad N+1\equiv 1,
\label{shift2}
\end{equation}
which is simply a consequence of the translation 
invariance of the transfer matrix (\ref{transfermatrix}). 
But also $v_1=\lambda$ in (\ref{eikj}) fixes 
 $e^{i k_1}=0$, what obliges to set
\begin{equation}
e^{\displaystyle{i\,k_1}}=-e^{\displaystyle{-i\,k_2}}=0,
\label{pair}
\end{equation}
as was done in \cite{B1}. This happens for all bound pairs that
we have obtained no matter the values of $N$ and $n$: it is 
simply a {\em fact} that for these states in this model 
\begin{equation}
e^{\displaystyle{i\,(k_1+k_2)}}=-1.
\label{prepair} 
\end{equation}
This condiction, together with the two identities in (\ref{pair})
mark how to work appropiately with bound pairs.
%$|1,2\ra-|2,3\ra+|3,4\ra-|1,4\ra$ when $N=4$ 

\section{n=3}
\label{n=3}

The trial function (\ref{definitionQ}) is now of the form
\begin{equation}
\tQ(z)=(z\,y-1)(z-y)(z-A)/z^{3/2},
\label{QN=6n=3}
\end{equation}
with $A$ a constant (numerical or depending on $y$) to be determined.
Substituting (\ref{QN=6n=3}) in (\ref{functionalrelation}), the
r.h.s. of this equation is {\em exactly} divided by $\tQ(z)$ in the
l.h.s. if and only if $A=0,-1,1$ or $A$ is the solution of a certain
polynomial whose coefficients depend only on $\Delta$.  The root
$A=0$ is not an admissible solution because (\ref{QN=6n=3}) 
has not the required expansion (\ref{definitionQ}); on the
contrary, roots $A=-1,1$ yield admissible functions $\tQ(z)$ because
the associated $\Lambda(v)$ by (\ref{functionalrelation}) are always
in the spectrum of the transfer matrix, as we have verified in
numerous experiments.  For example, the numbers
\begin{align}
\Lambda_+&\equiv 2\,a^3 b^3-a b c^2\,(a^2+a b+b^2)+c^4\,(a^2-a b+b^2),
\label{LambdapN=6n=3}\\
\Lambda_{-}&\equiv 2\,a^3 b^3-a b c^2\,(a^2-a b+b^2)-c^4\,(a^2+a b+b^2),
\label{LambdamN=6n=3}
\end{align} 
are eigenvalues of 
the $N=6$ transfer matrix for
arbitrary values of $a,b,c$.
The first is for $A=-1$, the second for $A=1$. 
We present some of these numerical tests in Table~1.
Regarding the situation in which $A$ is the solution of a certain 
polynonial, when $N=6$ such polynomial is
\begin{equation}
A^4+\left(8 \De^3-4 \De \right)
\,A^3+\left(20 \De^2-14\right)\,A^2
+\left(8 \De^3-4 \De \right)\,A+1=0,
\label{equationAn=3}
\end{equation}
but it has to de  discarded 
because none of the four roots of (\ref{equationAn=3}) is linked to
an eigenvalue of the transfer matrix for arbitrary $\Delta$
(it can be checked also with Table~1). There are only two $\tQ$'s 
(that is, two bound pairs in the block) and two eigenvalues.

\vspace{8pt}
\centerline{$N=6,\, n=3$}

\vspace{5pt} 
{\centerline{\footnotesize \begin{tabular}
{!{\vrule width 1pt}l!{\vrule width 1pt}l
!{\vrule width 1pt}l!{\vrule width 1pt}}
%\begin{tabularx}{.8\linewidth}{|X|X|X|}
\hline
 $z=\exp\,(i\,\pi/5)$& $z=\exp\,(i\, 2 \pi/5)$ &$z=64$ \\
 $y=\exp\,(i\,\pi/3)$  & $y=\exp\,(i\, 3 \pi/4)$ &$y=25 $  \\
 $\rho=i$   &$\rho=i $ &$\rho=1$ \\
$\De=-1/2$&$\De=\sqrt{2}/2=0.707\ldots$ &$\De= 
-313/25=-12.52$\\
\hline
&&$\qquad\qquad\times 10^{6}$\\\cline{3-3}
$0.779\,508$ &$2.028\, 398 $ &$ 24.899\, 897$ \\
$0.270\,722$ &$0.421\,472 $ &$10.523\, 295\,^{+} $\\
$0.168\,519$&$0.305\, 370$ &$8.909\,579$ \\
$0.168\,519\,^{+}$ &$0.074\, 622\, 2 $ &$ 3.719\,286 $ \\
$-0.105\,922$&$0.065\, 240\, 7 $ &$-3.724\,777$ \\
$-0.229\,696$ &$0.001\, 298\, 21\,^{+} $ &$-8.840\,959\,^{-}$ \\
$-0.465\,600\,^{-}$ &$-0.349\, 829\,^{-}$&$-10.587\,344$ \\
$-0.527\,034$ &$-0.448\, 262 $ &$-24.913\,779$\\
$0.265\,632  \pm 0.336\,454\,i$&$ 0.851\, 314\pm 0.535\, 651\,i$
&$ 12.449\,161\pm 21.567\,683\,i$\\
$0.086\, 642\, 0 \pm 0.196\, 887\,i$&$ 0.395\, 856\pm 0.293\, 548\,i$ 
&$4.891\,466\pm7.822\,856\,i$ \\
$0.084\, 086\, 1  \pm 0.159\, 490\,i$&$0.184\, 229\pm 0.002\, 347\, 62\,i$ 
&$4.750\,077\pm 8.912\,538\,i$ \\
$- 0.084\, 223\,2\pm 0.124\,642\,i$&$0.098\,987\,5\pm 0.098\, 842\, 6\,i $ 
&$ -4.756\,978 \pm 8.869\,031\,i$ \\
$-0.131\, 171 \pm 0.269\,132\,i$&$0.039\,950\,5\pm 0.112\, 727\,i $ 
&$-4.887\,688 \pm 7.863\,040\,i$ \\
$-0.176\, 703 \pm 0.299\, 057\,i$ &$0.003\,395\, 86\pm 0.451\, 325\,i$ 
&$-12.457\,140\pm 21.571\,004\,i$ \\
\hline
$a=\sin(\pi/15)=0.207\ldots$ & 
$a=\sin(7\pi/40)=0.522\ldots$& 
$a=39/80=0.4875$\\
$b=\sin(4\pi/15)=0.743\ldots$& $b=\sin(23\pi/40)=0.972\ldots$& 
$b=-1599/80=-19.9875$\\
$c=\sqrt{3}/2=0.866\ldots$&$c=\sqrt{2}/2$&$c=-312/25=-12.48$\\
\hline
\end{tabular}}
}

\vspace{4pt}

\renewcommand{\baselinestretch}{0.8}
\noindent\textsf{\footnotesize
Table~1.
In vertical are shown the
$20$ eigenvalues  
of the transfer matrix 
block $N=6, n=3$ for different values of $a,b,c$.
The eigenvalues are obtained by numerical diagonalization
of the matrix in (\ref{transfermatrix}), and each result approximated
to the number arrayed in the table with the rule of $5$. In all the 
examples we have fixed 
$z,y,\rho$, and $a,b,c$ are derived from them through (\ref{vlambda}).
The values marked 
with $+$ and $-$ coincide, no matter the number of digits 
of accuracy demanded in the computation, 
with the theoretical values 
(\ref{LambdapN=6n=3}), (\ref{LambdamN=6n=3}) obtained in this 
paper solving (\ref{functionalrelation}).   
In the third column it is necessary to multiply by
$10^6$ to obtain 
the correct eigenvalue. Notice that when $\Delta=-1/2$ the bound 
pair (\ref{vectorplus}) 
is degenerated and the transfer matrix has another 
linearly independent proper state
with the same eigenvalue $a^6+b\,^6$. This degeneration
happens for all values of $a,b,c$ 
and not only for the particular value
listed here.
}

\renewcommand{\baselinestretch}{1}

\vspace{5pt} 

The situation is the same for arbitrary $N$ even: there are only two bound 
pairs in the block and the generalization of 
(\ref{LambdapN=6n=3}) and (\ref{LambdamN=6n=3}) is
\begin{align}
\Lambda_{+}&= a^3 b^3\,(a^{N-6}+b^{N-6})
-a b c^2\,\left(a^{N-4}+b^{N-4}+a b\, \frac{a^{N-5}+b^{N-5}}{a+b}\right)
+c^4\,\left(\frac{a^{N-3}+b^{N-3}}{a+b}\right),
\nn\\
%\label{LambdapN=6n=3}\\
\Lambda_{-}&= a^3 b^3\,(a^{N-6}+b^{N-6})
-a b c^2\,\left(a^{N-4}+b^{N-4}-a b\,\frac{a^{N-5}-b^{N-5}}{a-b}\right)
-c^4\,\left(\frac{a^{N-3}-b^{N-3}}{a-b}\right),
\nn
%\label{LambdamN=6n=3}
\end{align} 
that correspond to 
\begin{equation}
\tQ^{+}(z)=(z\,y-1)(z-y)(z+1)/z^{3/2}\quad{\rm and}\quad
\tQ^{-}(z)=(z\,y-1)(z-y)(z-1)/z^{3/2},
\label{2Q}
\end{equation}
respectively. 
The quotients written in $\Lambda_{\pm}$ above are 
fictitious\footnote{I.e., introduced by the author to make the expressions 
compact} because the divisions can be performed exactly giving as result 
polynomials in $a,b,c$ with no denominators.

Each eigenvector of the transfer matrix has associated a given 
$\tQ(z)$, we now calculate as an example the eigenvector 
associated to $\tQ^{+}$ in (\ref{2Q}) for $N=6$ using 
Bethe ansatz.\footnote{
We insist on the words {\em Bethe ansatz} because
some authors refer to bound pair states as {\em non-Bethe states},
and they are Bethe states}
For such state the  product
\begin{equation}
e^{\displaystyle{i\,(k_1+k_2+k_3)}}=1,
\label{e123} 
\end{equation}
that can be justified in several manners: one, if the 
eigenvalue is known, (\ref{LambdapN=6n=3}) in this case, it is enough
to set $b=0$, $a=c$ in the eigenvalue.  The coefficient of $c^N$ is
precisely $e^{i\,(k_1+\cdots+k_n)}$ \cite{Baxterbook}; 
or two, evaluating 
$(-1)^n\tQ(zy^2)/\tQ(z)$ 
at the point $z=1/y$ \cite{B1}. 
This gives also such product. Since
the third zero of the function $\tQ^+$ is at $z_3=-1$, relation (\ref{eikj})
indicates that $e^{i\,k_3}=-1$, that substituted in (\ref{e123}) gives
the product $e^{i\,(k_1+k_2)}=-1$, something that seems to be 
shared by all bound pairs of the model as we remarked in 
(\ref{prepair}). For our pair holds again (\ref{pair}) what makes
that the factor $s_{21}$ vanishes according to (\ref{sij}). 
To obtain the correct bound pair state the rule is\footnote{We have 
taken this rule from \cite[Sect.~4]{B1}}:
calculate the $s_{ij}$ that do not vanish
(in the present case there are five of them) with (\ref{sij}), 
keeping only the dominant term as 
$e^{{i k_1}}$ goes to zero,
and calculate $s_{21}$ with (\ref{amplitude}). 
In this manner, instead of
writting $`s_{21}=0\!$' in the formulae, $s_{21}$ takes 
the expression that vanishes most rapidly as $e^{{i k_1}}$ goes to zero.
This expression is
\begin{equation}
s_{21}=2\,\De\,\l(1+2\,\De\r)\,e^{\displaystyle{i (N-1) k_1}},
\label{another-s21} 
\end{equation}
while 
\begin{alignat}{2}
s_{12}&=2\,\De\,e^{\displaystyle{-i k_1}},&\qquad
&s_{13}=1+2\,\De,\qquad s_{31}=1,\nn\\
s_{23}&=e^{\displaystyle{-i k_1}},&\qquad
&s_{32}=\l(1+2\,\De\r)\,e^{\displaystyle{-i k_1}}.\label{s31}
\end{alignat}
Note that the amplitudes obtained with (\ref{As})
after the substitution of 
(\ref{another-s21}) and 
(\ref{s31}) 
do satisfy exactly equations
(\ref{amplitude}), as expected. Take now $N=6$. 
Inserting the values (\ref{As})
into (\ref{Bethestate}) we find that 
for example, 
$f(1,2,3)=-2\,\De\,\l(1+2\,\De\r)^2/C$ and
$f(1,2,4)=2\,\De\,\l(1+2\,\De\r)\,e^{\displaystyle{-i k_1}}/C$.
In the case $N=6$ 
two more components are necessary to write the eigenvector, namely
\[
f(1,2,5)=-2\,\De\,\l(1+2\,\De\r)\,e^{\displaystyle{-i k_1}}/C,
\qquad 
f(1,3,5)=-6\,\De\,\l(1+2\,\De\r)/C,
\]
since the remaining components are deduced from these four 
with the generalization of property (\ref{shift2})
to the case  $n=3$.
Clearly $f(1,2,4), \,f(1,2,5)$ are the elements that 
grow most rapidly as 
$e^{{i k_1}}$ vanishes and the sensible choice here is to take $C$  
so that $f(1,2,4)=1$. The result 
is the right eigenvector associated to 
 $\tQ^+$ in (\ref{2Q})\footnote{For general 
$N$ the eigenvector is 
$|\psi\ra=\sum_{l=1}^N(|l,l+1,l+3\ra-|l,l+2,l+3\ra)$. The state 
that accompanies to $\tQ^-$ is 
$|\psi\ra=\sum_{l=1}^N (-1)^l (|l,l+1,l+3\ra+|l,l+2,l+3\ra)$. It has some 
similary with
(\ref{knowntoBethe}) but in the block $n=3$}
\begin{align}
|\psi\ra=&|1, 2, 4\ra +|2, 3, 5\ra+|3, 4, 6\ra+|1, 4, 5\ra
+ |2, 5, 6\ra + |1, 3, 6\ra 
\nn\\
&-|1, 2, 5\ra-|2, 3, 6\ra-|1, 3, 4\ra-|2, 4, 5\ra-|3, 5, 6\ra
-|1, 4, 6\ra,
\label{vectorplus} 
\end{align}
which coincides with the vector found in \cite[eq. (22)]{Sid} using
different methods.

\section{n=4 and n=5}
\label{n=4}

There is no problem in repeating the same steps as in $n=3$ to
deduce the number of bound pairs when $n=4$ or $n=5$. In fact
introducing
\begin{equation}
Q(z)=(z\,y-1)(z-y)(z^2+A\,z+B)/z^2
\label{Qn=4AB}
\end{equation}
into (\ref{functionalrelation}), it is possible to find 
constants $A$ and $B$ so that
the function
$\Lambda(v)$ is an eigenvalue 
of the transfer matrix block $n=4$ for
arbitrary $a,b,c$ activities. 
However, we follow a different method in
this section with the intention of 
obtaining a better trial function $Q$ not as general as in 
(\ref{Qn=4AB}): 
we solve directly Bethe ansatz equations (\ref{amplitude}) 
instead\footnote{Once $e^{{i k_3}},\ldots,e^{{i k_n}}$ 
are found solving Bethe equations, we use (\ref{eikj}) to write  $\tQ$ 
given by (\ref{definitionQ})}. The
equations are already solved for $e_1$ and $e_2$ (for brevity we will
use from now the notation $e_1$ to denote the number 
$e^{{ik_1}}$, $e_2$ to denote $e^{{i k_2}}$, and so on), 
since we know that
$e_1=0$, $e_2=-1/e_1$, with the product $e_1 e_2$ equal to $-1$
as a characteristic of
bound pairs. It remains to solve for  $e_3, e_4$ in the case 
$n=4$, and for $e_3, e_4, e_5$ in the case of $n=5$. And
when resolving the same care about $s_{ij}$ has to be taken that when the 
eigenfunction (\ref{vectorplus}) was constructed in the previous section: 
$s_{21}$ that vanishes
has to be evaluated with (\ref{amplitude}), taking then
the expression that vanishes most rapidly as $e_1$ goes to zero, and the 
remaining $s_{ij}$
with (\ref{sij}). With these remarks taken into consideration
the equations to solve are  
\begin{align}
e_3^{N-1}&=-\left(\frac{1-2\Delta e_3}{e_3-2\Delta}\right)
\left(\frac{1-2\Delta e_3+e_3 e_4}{1-2\Delta e_4+e_3 e_4}\right), 
\label{ba4-1}\\
e_4^{N-1}&=-\left(\frac{1-2\Delta e_4}{e_4-2\Delta}\right)
\left(\frac{1-2\Delta e_4+e_3 e_4}{1-2\Delta e_3+e_3 e_4}\right),
\quad N\ge 8,
\label{ba4-2}
\end{align}
in the block $n=4$, and 
\begin{align}
e_3^{N-1}&=\left(\frac{1-2\Delta e_3}{e_3-2\Delta}\right)
\left(\frac{1-2\Delta e_3+e_3 e_4}{1-2\Delta e_4+e_3 e_4}\right)
\left(\frac{1-2\Delta e_3+e_3 e_5}{1-2\Delta e_5+e_3 e_5}\right),
\label{ba5-1}\\
e_4^{N-1}&=\left(\frac{1-2\Delta e_4}{e_4-2\Delta}\right)
\left(\frac{1-2\Delta e_4+e_3 e_4}{1-2\Delta e_3+e_3 e_4}\right)
\left(\frac{1-2\Delta e_4+e_4 e_5}{1-2\Delta e_5+e_4 e_5}\right),
\quad N\ge 10
\label{ba5-2}\\
e_5^{N-1}&=\left(\frac{1-2\Delta e_5}{e_5-2\Delta}\right)
\left(\frac{1-2\Delta e_5+e_3 e_5}{1-2\Delta e_3+e_3 e_5}\right)
\left(\frac{1-2\Delta e_5+e_4 e_5}{1-2\Delta e_4+e_4 e_5}\right),
\label{ba5-3}
\end{align}
when $n=5$. Remember that $\Delta$ is given by (\ref{defDelta1}) and      
$N$ is an even number.

Consider the equations relative to $n=5$ for a moment.
Notice that if $(e_3,e_4,e_5)$ is a solution of equations
(\ref{ba5-1})-(\ref{ba5-3}) for given $N$ and 
$\Delta$\footnote{$\Delta$ fixed though arbitrary}, also
$(e_4,e_3,e_5)$, the interchange of $e_3$ with $e_4$, is a solution;
and also it is $(e_3,e_5,e_4)$. Equations
(\ref{ba5-1})-(\ref{ba5-3}) 
do not distinghish         
a solution from any of its permutations.
It is for this reason that two solutions are considered the same
if coincide up to permutations.

There is another relevant property of the equations: if
$(e_3,e_4,e_5)$ is a solution,
$\left({\frac{1}{e_3}},{\frac{1}{e_4}},{\frac{1}{e_5}}\right)$ is also
a solution for the same $N$ and $\Delta$. This feature brings
considerable insight into the resolution of
(\ref{ba5-1})-(\ref{ba5-3}).  For example, if $e_3$ is in the solution
so does $1/e_3$, as this property establishes, therefore $1/e_3$
is one of the numbers in $(e_3,e_4,e_5)$. If it is equal
to its inverse, $e_3$ is $1$ or $-1$, but if not, the inverse of $e_3$
has to be say, $e_4$, and thus $e_3 e_4=1$. 
%with none of the two numbers equal to $1$ or
%$-1$ now.  
The argument is
repeated with $e_4$ to conclude that $e_4$ is $1$ or $-1$ or the
inverse of $e_3$. Finally, it is the turn of $e_5$, that can be only
$\pm 1$ and not the inverse of any other number because there are no
more left numbers to be paired with. In conclusion: $(e_3,e_4,e_5)$
are $(1,1,1)$, $(-1,-1,-1)$ or $(e_3,e_4,\pm 1)$, with $e_3
e_4=1$. There are no more possibilities for arbitrary $\Delta$.  
Something similar happens
when $n=4$: the only solutions $(e_3,e_4)$ of (\ref{ba4-1}),
(\ref{ba4-2}) with $\Delta$ arbitrary
are $(1,-1)$ or the combinations $(e_3,e_4)$ that satisfy $e_3 e_4=1$. 
Obviously this is so because the two 
properties explained above, permutation and inversion,
hold for equations (\ref{ba4-1}), (\ref{ba4-2}) as 
well\footnote{Observe that for all bound pairs obtained so far the
product $e_1\cdots e_n=\pm 1$, something already mentioned in 
\cite{B1} and \cite{Sid}. The momentum of these states, 
the sum of the $k$'s, is therefore 
$0$ or $\pi$ (mod $2\pi$)}.
\begin{lemma}
\label{lemman=4}
$(n=4)$ The numbers $e_3, e_4$ given by equations 
(\ref{ba4-1}), (\ref{ba4-2}) subject to the
condition $e_3 e_4=1$, are the roots of the 
quadratic polynomial
\begin{equation}
x^2-(r+{1}/{r})\,x+1=0,
\label{polx4}
\end{equation}
where $r$ is, in turn, the solution of the polynomial of degree $N$
with coefficients fixed by $\Delta$ given by
\begin{equation}   
r^{N}-3\,\Delta r^{N-1} + 2\,\Delta^2 (r^{N-2}+r^2)-3\,\Delta r+1=0.
\label{pol4.N}
\end{equation}
\end{lemma} 
\begin{proof} Very simple.
Just substitute directly $e_3=r$,  
$e_4=1/r$ in (\ref{ba4-1}) and write the relation that results.
Zero solutions $r=0$ are not 
wanted\footnote{We want $e_3 e_4=1$ with $e_3$ and $e_4$ {\em finite} 
numbers. 
Therefore none of them vanishes. We do not want more {\em special}
objects like the pair $e_1 e_2=-1$ with $e_1=0$}. $\Box$
\end{proof} 
Surprisingly, the polynomial in (\ref{pol4.N}) has the same
coefficients when $N=8$, say, that when $N=100$, only that in this
case the coefficients are distributed according to a degree $100$.
Equality (\ref{pol4.N}) belongs to the class of {\em reciprocal equations}
\cite{Us} because the coefficient of $r^{N}$ is the same as the
independent term, the coefficient of $r^{N-1}$ the same as the
coefficient of $r$, and so on. If $R$ is a root of a reciprocal
equation, so it is its reciprocal $1/R$. This cannot be a surprise,
merely it is an expected consequence of the second property of the
Bethe equations remarked a few paragraphs above.
\begin{lemma}
\label{lemman=5}
$(n=5)$ The numbers $e_3, e_4, e_5$ given by equations 
(\ref{ba5-1})-(\ref{ba5-3}) with the
additional requirement $e_3 e_4=1$, $e_5=-1$, are the roots of the 
cubic polynomial
\begin{equation}
(x+1)\left(x^2+(r+{1}/{r})\,x+1\right)=0,
\label{polx5}
\end{equation}
where $r$ is the solution of (for simplicity we write the polynomial 
when $N=10$)
\begin{align}   
r^{10}+(5\,\Delta+2) r^9+& 2\,(2\,\Delta+1)^2 r^8+
2\,(2\,\Delta+1)(\Delta+1)^2 (r^7+r^6+r^5+r^4+r^3)\nn\\  
+& 2\,(2\,\Delta+1)^2 r^2+(5\,\Delta+2) r+1=0.
\label{pol5.10}
\end{align}
\end{lemma}
This is a reciprocal equation too. When $N$ is arbitrary, 
the polynomial that generalizes
(\ref{pol5.10}) is a polynomial of degree $N$: $r^{10}$, $r^{9}$, 
$r^{8}$ above change into
$r^{N}$,  $r^{N-1}$, $r^{N-2}$, respectively, and $r^{7}+\cdots+r^{3}$ 
into $r^{N-3}+\cdots+r^{3}$. Nothing else changes. With these directions 
we avoid to write the generalization explicitly.

{\leftskip=8pt \rightskip=8pt 

{\small 
\noindent
When the requirement is $e_3 e_4=1, e_5=1$,
the solution $(e_3,e_4,e_5)$ of equations 
(\ref{ba5-1})-(\ref{ba5-3}) is given by
$(x-1)\left(x^2+(r+{1}/{r})\,x+1\right)=0$, i.e., 
$e_3=-r, e_4=-1/r, e_5=1$, with $r$ the roots of the polynomial 
obtained changing $r$ by $-r$ and $\Delta$ by $-\Delta$ in
(\ref{pol5.10}). The polynomial thus obtained is 
generalized to other $N$'s with the directions 
explained in the previous 
lines.
\par
}
}

\begin{proof} 
The substitution of
$e_3e_4=1$ and $e_5=-1$ 
in (\ref{ba5-3}) gives no information because
the l.h.s. of (\ref{ba5-3}) reduces to a number and the r.h.s. to the
same number. However, substituted in (\ref{ba5-1}) (or in
(\ref{ba5-2})) is obtained a relation between the sum
$e_3+\ds{\frac{1}{e_3}}=e_4+\ds{\frac{1}{e_4}}=u$ and $\Delta$.  This
relation depends on $N$ and, for example, when $N=10$ is given by
\begin{align}   
{u}^{5}- \left(5\,\Delta+2 \right) {u}^{4}+
 &\left(8\,{\Delta}^{2}+8\,\Delta -3\right) {u}^{3}
-\left(4\,{\Delta}^{3}+10\,{\Delta}^{2}-12\,\Delta-6\right) {u}^{2}\nn\\
+ &\left(4\,{\Delta}^{3}-14\,{\Delta}^{2}-16\,\Delta+1\right) u
+2\, \left( 2\,\Delta-1 \right)  \left( {\Delta}^
{2}+3\,\Delta+1 \right)=0. 
\end{align}
It is hard to see any recurrence in this equation
but if $u$ is decomposed into a number and its inverse, i.e., as
$u=-(r+1/r)$, $r$ is a root of (\ref{pol5.10}), which 
is a much simpler equation than the previous one. 
The numbers $e_3=-r$, $e_4=-1/r$, $e_5=-1$, are therefore roots of 
(\ref{polx5}) with $r$ given by (\ref{pol5.10}) if $N=10$. $\Box$
\end{proof}

Now we count states. Starting with $n=4$, we have the state
characterized by 
$(e_1,e_2,e_3,e_4)=(e_1,-1/e_1,1,-1)$
obtained before Lemma~\ref{lemman=4}.
For this state $e_1 e_2 e_3 e_4=1$, and $\tQ$ and $\Lambda$ are given by
\begin{align}
&\tQ(z)=(z\,y-1)(z-y)(z^2-1)/z^2,
\label{Q1Nn=4}\\
&\La={a}^{4}{b}^{4}\left(a^{N-8}+b^{N-8}\right) 
-{a}^{2}{b}^{2}{c}^{2}\left( {a}^{N-6}+{b}^{N-6}-
2\, {a}^{2}{b}^{2}\,\frac { {a}^{N-
8}-{b}^{N-8} }{{a}^{2}-{b}^{2}} \right)
\nn\\ 
&\qquad\qquad-3\,{a}^{2}{b}^{2}{c}^{4}\left(
\frac {{a}^{N-6}-{b}^{N-6}}{{a}^{2}-{b}^{2}}\right )
+c^6\left(\frac {{a}^{N-4}-{b}^{N-4}}{{a}^{2}-{b}^{2}}\right ),
\quad N\ge 8
\label{L1Nn=4}
\end{align}
%\[
%2\,a^4 b^4-a^2 b^2 c^2\,(a^2+b^2)-3\,a^2 b^2 c^4+c^6\,(a^2+b^2),
%\]
as deduced from (\ref{eikj}), (\ref{definitionQ}) and the
relation (\ref{functionalrelation}). As in $\Lambda_{\pm}$ obtained in
Sect.~\ref{n=3}, the quotients in (\ref{L1Nn=4}) are artificial, and
the divisions can be performed exactly giving for $\Lambda$ an
homogeneous expression of order $N$ in $a,b,c$ with constant
coefficients. Regarding the solution $(e_1,-1/e_1,r,1/r)$ of
Lemma~\ref{lemman=4}, notice that since 
the roots of (\ref{pol4.N}) are single or at most double\footnote{
%There are roots of multiplicity two only for $\Delta=\pm 1/2,\pm 1$.
The discriminant of (\ref{pol4.N}) in $r$ vanishes only 
for $\Delta=\pm 1/2,\pm 1$, thus indicating multiplicity 
of the roots $r$ more than $1$ only for these values.
Why for these values? Notice that the bilinear transformation 
$e_3\to\frac{1-2\Delta e_3}{e_3-2\Delta}$ in the r.h.s. of
(\ref{ba4-1}) (and in the r.h.s. of (\ref{ba4-2}) for $e_4$) 
collapses to a constant when $\Delta=\pm 1/2$ instead of being
a one-to-one mapping. This justifies the multiplicities at
$\Delta=\pm 1/2$. A similar reason happens when  $e_3 e_4=1$ and 
$\Delta=\pm 1$ to the second factor in the r.h.s.  
of equations (\ref{ba4-1}), (\ref{ba4-2})},
there are $N/2$ different solutions because of the reciprocity of  
(\ref{polx4}) and (\ref{pol4.N}). 
For these $N/2$ solutions (i.e., states) $e_1 e_2 e_3 e_4=-1$, and 
$\tQ$ is given by
%\begin{equation}
%Q(z)=(z\,y-1)(z-y)\left(z^2-\left(t+\frac{1}{t}\right)\,z+1\right)/z^2,
%\label{Qn=4t1/t}
%\end{equation}
\begin{equation}
Q(z)=(z\,y-1)(z-y)\left(z^2-\left(t+1/t\right)\,z+1\right)/z^2,
\label{Qn=4t1/t}
\end{equation}
with 
\begin{equation}
t+\frac{1}{t}=-\frac{2\,\Delta\, (r+1/r)-4}{r+1/r-2\,\Delta},\qquad 
\Delta\neq \pm 1.
\label{t1/t}
\end{equation}
The number $\Lambda(v)$ is obtained 
inserting (\ref{Qn=4t1/t}) and (\ref{t1/t}) into
(\ref{functionalrelation}).
This result shows also that (\ref{Q1Nn=4}) and (\ref{Qn=4t1/t}) are
more accurate trial functions to solve (\ref{functionalrelation}) 
than the general (\ref{Qn=4AB}). Contrary to what we
have done along this paper, we do not write
the function $\Lambda(v)$ associated to
(\ref{Qn=4t1/t}) and (\ref{t1/t}) for general $N$, 
%(the interested reader
%will not have problem to obtain it using (\ref{functionalrelation})),
but we write it when $N=8$, which is 
\begin{equation}
\La(v)=2\,a^4 b^4+c^2 \left( 2\,\lambda_2\, a^3 b^3
-\lambda_3\, a^2 b^2 \,(a^2+b^2)
-2\, \lambda_1\, a b\,(a^4+b^4-a^2 b^2)-a^6-b^6\right),
\label{LN=8n=4e1e2e3e4=-1}
\end{equation}
with $\lambda_1, \lambda_2, \lambda_3$ certain numbers depending on 
$\Delta$ that we do not
specify.  The object to present (\ref{LN=8n=4e1e2e3e4=-1}) is to
comment about the excluded cases $\Delta=\pm 1$ pointed in
(\ref{t1/t}). We have excluded these two points for mathematical
reasons only. Let us fix $\Delta=1$ (we center the discussion in this
value because the polynomial (\ref{pol4.N}) indicates that the
situation when $\Delta=-1$ is the same just negating $r$). Subtituting
$\Delta=1$ in (\ref{t1/t}), the r.h.s.  reduces either to the constant
$-2$ or to the indetermination $0/0$\footnote{$r=1$ 
is solution of (\ref{pol4.N}) when $\Delta=1$}: which
is then the function (\ref{Qn=4t1/t}) and how many of them can one
write when $\Delta=1$?  We wont be
more explicit in
this point now, however we want to convince the reader 
that for $N=8, \Delta=1$ there are four (eventually $N/2$ for general
$N$, if things go as they shall) 
bound pair states with $e_1 e_2 e_3 e_4=-1$: 
we have just constructed the states 
(\ref{Bethestate}) with (\ref{As}), (\ref{sij})
and (\ref{amplitude}) imposing the conditions
(\ref{pair}) and (\ref{prepair}); we have obtained 
exactly four states,
and have checked (diagonalizing numerically the matrix block)
that they are eigenvectos of the
transfer matrix (\ref{transfermatrix}) when $N=8, n=4$.
The associated eigenvalues are precisely
(\ref{LN=8n=4e1e2e3e4=-1}) with
$\lambda_1=0, -3.69963,-1.76088,0.460505$\footnote{Approximated 
to the nearest six digit number the last three 
data}, and 
$\lambda_2, \lambda_3$ given in terms of
$\lambda_1$ by
\begin{equation}
\lambda_2=\frac{2-3\, \lambda_1^2- 4\,\lambda_1}{2 +\lambda_1}, 
\qquad \lambda_3=2\,\lambda_1^2+2 \lambda_1-1,\qquad \De=1.
\label{lambda23}
\end{equation}

\noindent
In conclusion, for each real value of $\Delta$ in   
the vertex model,
there are  $N/2+1$ bound pair states in the $n=4$ block of the 
$N$-site transfer matrix. The number of such states is 
correct\footnote{In
reference \cite{B1} were found $5$ states when $N=8, n=4$, as we
mentioned in Sect.~\ref{theproblem}. Our result agrees with that number} 
because exact diagonalization of the
block corroborates it: our numerical experiments carried up to
$N=12$ with different but arbitrary 
values of the activities $a,b,c$ confirm
that the numbers $\Lambda(v)$ obtained substituting 
$Q$ by (\ref{Qn=4t1/t}) with (\ref{t1/t}) and (\ref{pol4.N}) into
(\ref{functionalrelation}) 
are true
eigenvalues of the transfer matrix. The number 
(\ref{L1Nn=4}) is also an eigenvalue.
We have no reason then
to doubt that they are eigenvalues for general $N$ as well. 
The author thus admits the number $N/2+1$ as absolutely right. 

\vspace{10pt}

For $n=5$, we count a total of $N$ bound pairs. This is so because 
the solutions 
$(e_3,e_4,e_5)=(1,1,1), (-1,-1,-1)$ 
of equations (\ref{ba5-1})-(\ref{ba5-3})\footnote{We mentioned
these solutions in the paragraph
before Lemma~\ref{lemman=4}} do not afford eigenvalues of the transfer 
matrix for $\Delta$ generic. 
We noticed this fact from our numerical tests carried with 
different values of $a,b,c$ and $N=10,12$:
the numbers
$\Lambda$ obtained with (\ref{functionalrelation})
and $Q$ as in (\ref{definitionQ}) with zeros at 
$z_1=y, z_2=1/y, z_3=z_4=z_5=\pm 1$ and $y$ arbitrary, 
do not correspond to 
eigenvalues of the transfer matrix\footnote{The reason 
is that the states derived from these 
solutions proceeding as in Sections~\ref{BA} and~\ref{n=4}  
are the zero vector}. Unlike this,
the solutions in Lemma~\ref{lemman=5} that satisfy $e_3e_4e_5=-1$ 
afford $N/2$ bound pairs for each $\Delta$, 
and the solutions that satisfy $e_3e_4e_5=1$ 
afford another $N/2$ bound pairs (even for $\Delta=\pm 1$ in both cases). 
The corresponding numbers $\Lambda$ were checked numerically. 
These eigenvalues are obtained with 
\begin{equation}
\tQ(z)=(z\,y-1)(z-y)\left(z^2-\left(t+1/t\right)\,z+1\right)
(z\pm 1)/z^{5/2},
\label{Qn=5,pm}
\end{equation}
the plus sign in $\pm$ is for $e_1e_2e_3e_4e_5=1$ 
(i.e., $e_3e_4e_5=-1$), the minus sign for 
$e_1e_2e_3e_4e_5=-1$. In both functions written in
(\ref{Qn=5,pm})
\begin{equation}
t+\frac{1}{t}=-\frac{2\,\Delta\, (r+1/r)+4}{r+1/r+2\,\Delta},\qquad 
\Delta\neq \pm 1,
\end{equation}
but $r$ is the root of different polynomials, as stated 
in Lemma~\ref{lemman=5}. 

We write an example for $N=10$ and $\Delta=-1/2$, with the choice
of $z,y,\rho$ as
in the left column of Table~1. After diagonalizing numerically the
blocks $n=4,5$ of the transfer matrix, the eigenvalues corresponding
to bound pairs (we have recognized
them because they match exactly our predicted values) 
approximated to the nearest six digit number are\footnote{An 
interesting question is if they can be recognized in another manner}: 
%\vspace{8pt}
%\centerline{$N=10$}

\vspace{5pt} 
{\centerline
{\footnotesize 
%\begin{tabular}{|ll|llll|}
\begin{tabular}{|ll!{\vrule width 1pt}llll|}
\hline
$n=4$&[deg]&\multicolumn{1}{l}{$n=5$}&\multicolumn{3}{l|}{[deg]}
\\
\hline
\phantom{m} $0.111\,240\,^{+}$&$[2]$&$0.223\,385\,^{+}$&$[2]$&\qquad
$-0.065\,746\,4\,^{-}$&$[2]\;\,(r=2)$\\
$ - 0.040\,142\,3\,^{-}$&$[2]\;\;(r=-1)$&$0.138\,383\,^{+}$&$[2]$&\qquad
$-0.086\,922\,0\,^{-}$&$[2]\;\;(r=1)$\\
$ - 0.047\,993\,4\,^{-}$&$[1]$&$0.074\,747\,8\,^{+}$&$[2]$&\qquad
$-0.116\,844\,^{-}$&$[1]$\\
$- 0.080\,403\, 8\,^{-}$&$[1]$&$0.043\,830\, 8\,^{+}$&$[2]$&\qquad
$-0.226\,567\,^{-}$&$[1]$\\
$ - 0.158\,869\,^{-}$&$[1]$&$0.032\, 535\, 7\,^{+}$&$[2]$&\qquad
$-0.463\,488\,^{-}$&$[1]$\\
$ - 0.280\,662\,^{-}$&$[1]$&&&&\\
\hline
\end{tabular}
}}

\vspace{4pt}

\renewcommand{\baselinestretch}{0.8}
\noindent\textsf{\footnotesize
Table~2. Each eigenvalue listed is followed by a 
sign $+$ or $-$: the sign 
$+$ indicates that  $e_1e_2e_3e_4=1$ (or 
that $e_1e_2e_3e_4e_5=1$ if $n=5$), the sign $-$ that the product
of the Bethe roots is $-1$. 
The degeneration of the eigenvalue
is [deg]. In the column corresponding to 
$n=4$, the number $0.111\,240$ coincides with (\ref{L1Nn=4}), and the
remaining five values agree with the theoretical $\Lambda$ 
obtained inserting (\ref{Qn=4t1/t}) 
and (\ref{t1/t}) into
(\ref{functionalrelation}). The eigenvalue that 
corresponds to $r=-1$, remember that in this column
$r$ is a solution of (\ref{pol4.N}), is degenerated. 
This degeneration is not a surprise, 
because it is a case in which two Bethe roots coincide ($e_3=e_4=-1$), 
and when it is true
that the eigenvector associated to such cases is usually the zero
vector, when $\Delta=-1/2$ it is not. Regarding the list when $n=5$, 
the values with a $+$ correspond to solutions $r$ of (\ref{pol5.10}),
and the values with a $-$ to solutions $r$ of the polynomial 
that is obtained changing in (\ref{pol5.10}) the variables   
$r,\Delta$ by $-r,-\Delta$. Totally expected is the degeneration of
the eigenvalue $-0.086\,922\,0$ since $e_3=e_4=-1,e_5=1$. But the
degeneration of $-0.065\,746\,4$ which happens for $e_3=-2,
e_4=-1/2, e_5=-1$ is less expected.   
}

\renewcommand{\baselinestretch}{1}

The last comment of the paper: the numerators of (\ref{Qn=2}), 
(\ref{2Q}), (\ref{Q1Nn=4}), (\ref{Qn=4t1/t}) and 
(\ref{Qn=5,pm}) are polynomials in
in $z$ with a reciprocal 
property, that is  
if $R$ is a solution, so it is $1/R$.
When looking for other $Q's$ in $n=7$ (say) one has to restrict to
numerators with this property.

% similartity of 8.7 and 8.14

\section*{Acknowledgments}

I am pleased to thank Prof.~J.~Shiraishi and the organizers of the
RIMS 2004 Symposium, {\sl Recent progress in Solvable Lattice Models},
held in Kyoto for allowing me to expose these ideas. In my work I am
grateful to G.~\'Alvarez Galindo for resolving some of my doubts.  But
to whom I feel inevitably grateful every day is to Pepe Aranda: 
seventy times seven I have knocked on his door asking about
polynomials, roots and other matters of Calculus, and seventy 
times seven he has received me without ever showing the slightest 
unwelcome gesture in his face or manners that prevented me from 
knocking on his door again.

This work is financially supported by the Ministerio de Educaci\'on 
y Ciencia of Spain through grant No. BFM2002-00950.

\end{document}